\documentclass[twocolumn,showpacs,preprintnumbers,amsmath,amssymb,floatfix,prc,
superscriptaddress]{revtex4}
\usepackage{graphicx}
\usepackage{dcolumn}
\usepackage{bm}

\def\beq{\begin{equation}}
\def\eeq{\end{equation}}
\def\be{\begin{eqnarray}}
\def\ee{\end{eqnarray}}
\def\hs{\hat{s}}
\def\htm{\hat{t}}
\def\hu{\hat{u}}

\newcommand{\f}[2]{\frac{#1}{#2}}
\newcommand{\dd}  { {\textrm d}}
\newcommand{\ba}{\begin{eqnarray*}}
\newcommand{\ea}{\end{eqnarray*}}

\usepackage{color}

\newcommand{\lsim}{
 \mathrel{\setbox0=\hbox{$<$}\raise0.6ex\copy0\kern-\wd0
 \lower0.65ex\hbox{$\sim$}}}

\newcommand{\gsim}{
 \mathrel{\setbox0=\hbox{$>$}\raise0.6ex\copy0\kern-\wd0
 \lower0.65ex\hbox{$\sim$}}}

\begin{document}
\title{Charged hadrons and nuclear parton distributions in
  p(d)A collisions}
\author{Adeola Adeluyi}
\affiliation{Department of Physics and Astronomy \\
Texas A\&M University-Commerce, Commerce, TX 75429, USA}
\author{Trang Nguyen}
\affiliation{Center for Nuclear Research, Department of Physics \\
Kent State University, Kent, OH 44242, USA}
\author{Bao-An Li}
\affiliation{Department of Physics and Astronomy \\
Texas A\&M University-Commerce, Commerce, TX 75429, USA}

\date{\today}
\begin{abstract}
Nuclear gluon modifications are the least constrained component of
current global fits to nuclear parton distributions, due to the
inadequate constraining power of presently available experimental data from
nuclear deep inelastic scattering and nuclear Drell-Yan lepton-pair
production. A recent advance is the use of observables from
relativistic nucleus-nucleus collisions to supplement the data pool 
for global fits. It is thus of interest to investigate the sensitivity
of various experimental observables to different strengths of
nuclear gluon modifications from large to small Bjorken $x$.
In this work we utilize three recent global fits with different gluon strengths 
to investigate the sensitivity of three observables: nuclear
modification factor, pseudorapidity asymmetry, and charge ratio.
We observe that both nuclear modification factor and pseudorapidity
asymmetry are quite sensitive to the strength of gluon modifications in
a wide pseudorapidity interval. The sensitivity is greatly enhanced at LHC
(Large Hadron Collider) energies relative to that at RHIC 
(Relativistic Heavy Ion Collider). The charge ratio
is mildly sensitive only at large Bjorken x. Thus measurement of these
observables in proton-lead collisions at the LHC affords the
potential to further constrain gluon modifications in global fits. 
\end{abstract}
\maketitle
\vspace{1cm}
%
%
\section{Introduction}
\label{intro}

Nuclear parton distribution functions (nPDFs) are an important
ingredient in the theoretical description of nucleus-nucleus
collisions using the framework of perturbative Quantum Chromodynamics (pQCD). 
Although the scale evolution of the nuclear parton distributions can be
described perturbatively, nPDFs are intrinsically
non-perturbative. This non-perturbative nature makes their theoretical
determination from QCD hitherto intractable. Therefore, as obtains in
the case of nucleon parton distribution functions (PDFs), the nPDfs
are currently most efficiently determined from global fits to
experimental data. Earlier global analyses 
\cite{Eskola:1998df,deFlorian:2003qf,Shad_HKN,Hirai:2007sx} relied heavily on
fixed-target nuclear deep-inelastic scattering
(DIS) and Drell-Yan (DY) lepton-pair production, with the attendant
low precision and almost total lack of constraints on nuclear gluon
distributions. The use of an extended data set, incorporating data on
inclusive hadron production in deuteron-gold collisions, has been
pioneered in \cite{Eskola:2008ca,Eskola:2009uj}, with better constraints on gluon
modifications. Despite all these advances the nuclear gluon
distribution is still the least constrained aspect of global fits to 
nPDFs.

Thus questions related to the distributions of
partons (in particular gluons) in nucleons and nuclei remain of
current interest, both theoretically and experimentally. 
Of particular importance are the 
marked suppression of the nuclear modification factor at forward
rapidities and the issue of gluon saturation (see, for example
\cite{Gelis:2010nm,Kharzeev:2002ei}).
The Large Hadron Collider (LHC), even more than the Relativistic Heavy
Ion Collider (RHIC), affords the opportunity to
study several important aspects of these nuclear parton distributions  
in a center-of-mass energy and transverse-momentum domain where pQCD 
is expected to work well. For precision tests of pQCD applied to
nucleus-nucleus collisions and also for various other applications it
is important to gain control over the uncertainties in the elucidation
of nuclear gluon distributions.
It is therefore beneficial to investigate the sensitivity of various
observables to varying strengths of gluon modifications in different
Bjorken-$x$ regimes to further
enhance the constraining ability of global fits. 

In general the description of nuclear collisions based on pQCD is a complicated 
task. Much of the complication derives from the intrinsically complex nature 
of the nuclear environment in collisions of heavy nuclei. In addition to the 
initial-state modification of the PDFs, the cross section of high-$p_T$ 
hadron production is influenced by final-state effects (such as jet 
energy loss) and a complicated geometry. To better understand the physics of 
pQCD in the nuclear environment, it is highly desirable to disentangle the 
different nuclear phenomena affecting high-$p_T$ ($p_T \gsim 1.5$ GeV) hadron 
production. In this respect, asymmetric light-on-heavy nuclear
collisions, such as p(d)A collisions offer the distinct advantage of
absence of final-state effects. In fact, one of the physics goals of
the Run 8 at RHIC was to provide 
a high-statistics ``cold'' nuclear matter data set to establish a definitive baseline 
for ``hot'' nuclear matter, created in gold-gold collisions. The benchmark role of the
``the deuteron-gold control experiment'' for e.g. energy-loss studies has often been 
emphasized~\cite{Gyulassy:2004jj,Hemmick:2004jc}. Proton-nucleus and
deuteron-nucleus reactions have also been used to study the 
Cronin-effect~\cite{Cron75,Antr79}. 
These asymmetric systems also offer unique information about the 
underlying dynamics not available in symmetric proton-proton or 
nucleus-nucleus collision systems. For example, a new feature,
offered by nonidentical colliding beams like d+Au, is the pseudorapidity asymmetry,
examined in some detail by the STAR
collaboration~\cite{Abelev:2006pp}. In view of the importance of
asymmetric collisions, we anticipate similar pPb program at the LHC.

The strategy adopted in this study is as follows: we consider three
global nuclear parton distribution sets \cite{Hirai:2007sx,Eskola:2008ca,Eskola:2009uj} with 
different strengths of the nuclear gluon
modifications. We are mainly interested in the sensitivity of
observables to gluon modifications; therefore we consider these 
nPDF sets simply as sources of differing severity of gluon modifications.
Further details about this aspect of the study are given in the next section. 
In order to avoid issues related to final-state effects
which are present in nucleus-nucleus collisions,
we limit our calculations to charged hadron production in both 
deuteron-gold collisions at RHIC and
proton-lead collisions at the LHC. Utilizing this framework, we
investigate the sensitivity of three
basic quantities directly observable experimentally: the nuclear
modification factor, the pseudorapidity asymmetry, and the charge
ratio, to different strengths of the gluon modifications. The
calculated results are compared with available experimental data:
nuclear modification factor for charged hadrons from the BRAHMS
Collaboration~\cite{Arsene:2004ux}, and pseudorapidity asymmetry from the STAR
Collaboration~\cite{Abelev:2006pp} respectively. 

The paper is organized as follows: in Sec.~\ref{col} we review the basic 
formalism of the collinearly-factorized pQCD-improved parton model as applied to
nucleus-nucleus collisions. 
This section also includes the definition of the minimum bias nuclear
modification factor, pseudorapidity asymmetry, and charge ratio. 
We present the results of our calculations for these quantities in
Sec.~\ref{res}. Our conclusion is contained in Sec.~\ref{concl}.

\section{Theoretical framework and experimental observables}
\label{col}

\subsection{Collinear factorization}
\label{partmd}
The minimum bias invariant cross section for the production of 
final hadron $h$ from the collision of nucleus $A$ (with mass number
A) and nucleus $B$ (with mass number B)
($A+B \!\to\! h+X$), can be written, in collinearly factorized pQCD, as
\begin{eqnarray}\label{eq:pdA}
E_h\f{\dd ^3 \sigma_{AB}^{h}}{\dd^3 p} = AB
\sum_{\!\!abcd}\!\int\!\! \dd{x_a} \dd{x_b}
\dd{z_c} \,
F_{\!a/A}(x_a,Q^2) 
\nonumber \\ 
\times \,\ F_{\!b/B}(x_b,Q^2)\
\f{\dd\sigma(ab\!\to\!cd)}{\dd\htm}\,
\nonumber \\
\times \,\ \frac{D_{h/c}({z_c},\!Q_f^2)}{\pi z_c^2} \,
\hs \, \delta(\hs\!+\!\htm\!+\!\hu)   \,\,  
\end{eqnarray}
Here $x_a$ and $x_b$ are parton momentum fractions in $A$ and $B$, respectively, 
and $z_c$ is the fraction of the parton momentum carried by the final-state 
hadron~$h$.
The factorization and fragmentation scales are denoted as $Q$ and $Q_f$, respectively. 
As usual, $\hs$, $\htm$, and $\hu$ refer to the partonic Mandelstam variables, 
and the massless parton approximation is used in this study as
expressed in the delta function. 
The quantity $\dd\sigma(ab\!\to\!cd)/\dd\htm$ in Eq.~(\ref{eq:pdA}) represents the 
perturbatively calculable partonic cross section, and $D_{h/c}(z_c,\!Q_f^2)$ stands 
for the fragmentation function of parton $c$ to produce hadron $h$,
evaluated at momentum fraction $z_c$ and fragmentation scale
$Q_f$. For a recent discussion on the issue of collinear factorization
in nucleus-nucleus collision see \cite{QuirogaArias:2010wh}. 

The homogeneous nuclear parton distribution functions (nPDFs)
$F_{a/A}(x,Q^2)$, for a nucleus with $N$ neutrons and $Z=A-N$
protons, with $A$ the mass number, can in general be written as
\begin{eqnarray}\label{eq:npdf_A}
F_{\!a/A}(x,Q^2) = \frac{Z}{A}f_{a/p/A}(x,Q^2) + \frac{N}{A}f_{a/n/A}(x,Q^2)
\nonumber \\ 
= (1-\omega)f_{a/p/A}(x,Q^2) + \omega f_{a/n/A}(x,Q^2)
\end{eqnarray}
with $\omega = N/A$ and $0 \le \omega \le 1$.
The function $f_{a/p/A}(x,Q^2)$ denotes the parton distribution for parton $a$ in
a bound proton in nucleus $A$ and $f_{a/n/A}(x,Q^2)$ the distribution
in a bound neutron respectively, and are related, for valence quarks, by isospin symmetry. 
They are 
expressible as convolutions of free nucleon parton distribution functions (PDFs)
$f_{a/N}(x,Q^2)$ and a shadowing function ${\cal S}_{a/A}(x,Q^2)$ which encodes the 
nuclear modifications of parton distributions in nucleus $A$. 
We employ three different nuclear parton distributions in this study:
the HKN07~\cite{Hirai:2007sx} ("weak" gluon modifications),  
the EPS08~\cite{Eskola:2008ca} ("strong" gluon
modifications), the EPS09~\cite{Eskola:2009uj} ("moderate" gluon
modifications) sets. Note that for HKN07 and EPS09 the qualifiers 
"weak" and "moderate" respectively are with respect to the central
fits; inclusion of uncertainties generates considerable spread in the 
relative measure of gluon modifications. 
For consistency reasons we use the
underlying nucleon PDFs employed in the HKN07 nPDFs, namely
MRS98~\cite{Martin:1998sq}. The results are quite insensitive to the choice
of nucleon PDFs in general.
For the final hadron fragmentation we utilize the fragmentation functions
in the DSS07 set~\cite{de Florian:2007hc}. The factorization scale as well
as the fragmentation scale are set to evolve with the transverse
momentum of the outgoing hadron, according to $Q = Q_f =  p_T$.
Since pQCD calculations are generally not reliable at very low $p_T$,
the lower limit on our calculations has been set to $p_T = 1.5$ GeV/c. 
With our scale choice, this 
corresponds to a minimum $Q^2$ ($Q^{2}_{f}$) of $2.25$ GeV$^2$, while  
EPS08, EPS09, and HKN07 give $1.69, 1.69$, and $1.0$ GeV$^2$ for minimum $Q^2$, 
respectively. The minimum $Q^{2}_{f}$ for the DSS fragmentation 
functions is given as $1.0$ GeV$^2$.

It should be noted that whereas both EPS09 and HKN07 are available in
both leading order (LO) and next-to-leading order (NLO)
parametrizations, EPS08 is available only in leading order. Our
calculations are therefore to leading order. Since we consider
charged hadrons, we need charge-separated fragmentation
functions. These fragmentation functions are also available at LO and
NLO, with the NLO set adjudged more accurate \cite{de Florian:2007hc}. We
have performed calculations using both, and the results presented
here are for the NLO set. The implication of using NLO
fragmentation functions in a LO calculation is treated in
\cite{Albino:2008fy}.

A comprehensive treatment of the uncertainties on calculated 
results has not been undertaken in the present study.
The overall uncertainties on the calculated results obviously depend on
three components: the nPDFs, truncation of the perturbative series for
the hard partonic cross sections, and the charged hadron fragmentation
functions.
Although both  EPS09 and HKN07 have facilities for estimating the
uncertainties in the nPDFs, we have not included these uncertainties
in our calculations. We have used the central fits in these two
routines without recourse to their associated uncertainties. 
While the inclusion of uncertainties
may be important for comparing the relative efficacy of different
nPDFs, the focus here is on the sensitivity of the observables 
to different gluon modification scenarios. It is thus more transparent
for the purpose at hand to use the central fits in HKN07 and EPS09,
but bearing in mind the potential overlap of calculated results  
with the inclusion of uncertainties in these nPDFs. 
The calculated quantities in the present study are ratios,
therefore truncation errors largely cancel (for a comparison of
LO and NLO nuclear modification factor, see, for instance,
\cite{deFlorian:2003qf}). The errors arising from fragmentation
functions are basically unknown.  
%
\subsection{Nuclear modification factor}
\label{nucmod}

The minimum bias nuclear modification factor compares, as a ratio, spectra 
of particles produced in nuclear collisions to a hypothetical scenario 
in which the nuclear collision is assumed to be a superposition of the 
appropriate number of proton-proton collisions. The ratio can be 
defined as a function of $p_T$
for any produced hadron species $h$ at any pseudorapidity $\eta$:
\begin{equation}
R^h_{AB}(p_T, \eta) = \frac{1}{AB} \cdot
\frac{E_h \dd^3\sigma_{AB}^{h}/\dd^3 p}
{E_h \dd^3\sigma_{pp}^{h}/\dd^3 p}   \,\, 
\label{rAB}
\end{equation}
Nuclear effects manifest themselves in $R^h_{AB}(p_T)$ values greater or 
smaller than unity, representing enhancement or suppression, respectively, 
relative to the $pp$ reference.

The determination of the nuclear modification factor involves the
cross section for proton-proton collisions in the denominator, and is
thus sensitive to the precision of the knowledge of the $pp$ cross
section. The other two observables considered, the pseudorapidity
asymmetry and charge ratio, are independent of a baseline $pp$ cross
section and therefore do not suffer from this constraint.
\subsection{Pseudorapidity asymmetry}
\label{rapasym}

In general, as the mechanisms for hadron production in p(d)A collisions may be different
at forward rapidities (p(d) side) and backward rapidities (A side),
it is of interest to study ratios of particle yields between a given 
pseudorapidity value (interval) and its negative in these collisions. 
Thus the pseudorapidity asymmetry, $Y^h_{AB}$, is defined as the ratio of
the yield in
the backward (``target-side'', negative rapidities) region relative to the yield
in the forward (``projectile-side'', positive rapidities) region:
\begin{equation}
Y^h_{AB}(p_T) = \left. E_h\f{\dd ^3 \sigma_{AB}^{h}}{\dd^3 p} \right|_{-\eta} 
 \left/ 
\left. E_h\f{\dd ^3 \sigma_{AB}^{h}}{\dd^3 p} \right|_{\eta} \right. \,\, .
\label{yAB}
\end{equation}

The meaning of the pseudorapidity asymmetry is clear: a ratio greater
than unity implies more yield in the backward (A side) region relative 
to the forward (p(d)) region while a ratio less than unity implies
the converse. Since the nPDFs are an important ingredient in hadron
production, the study of pseudorapidity asymmetry can offer valuable
information in further constraining the nuclear parton distributions.
As remarked above, the pseudorapidity asymmetry is independent of a
reference $pp$ cross section and thus more "self-contained"; it is
also more global in the sense that it encapsulates information on
nPDFs at different Bjorken-$x$ values. For conciseness, we will refer
to $Y^h_{AB} > 1$ as positive asymmetry and $Y^h_{AB} < 1$ as negative asymmetry.

\subsection{Charge ratio and rapidity asymmetry of charge ratio}
\label{chrat}

Let us define the charge ratio as the ratio of the yield of negatively
charged hadrons to the yield of the positively charged hadrons. Then 
we have
\begin{equation}
Z^h_{AB}(p_T, \eta) = \left. E_h\f{\dd ^3 \sigma_{AB}^{h}}{\dd^3 p}
\right|_{h = h^-} 
 \left/ 
\left. E_h\f{\dd ^3 \sigma_{AB}^{h}}{\dd^3 p} \right|_{h = h^+} \right. \,\, .
\label{chrAB}
\end{equation}

The charge ratio is expected to be more sensitive to the fragmentation
functions than to the nuclear parton distributions, especially as one
moves toward forward (large positive) rapidities. Nevertheless, in the
backward direction (negative rapaidites) and even near midrapidity, the
charge ratio shows some sensitivity to nuclear parton
distributions. This rapidity dependence is conveniently displayed  
in terms of a 
pseudorapidity asymmetry of the charge ratio, $Y^{AB}_{Z}(p_T)$, defined
as the ratio of the charge ratio at negative rapidities relative to
that at positive rapidities. Technically this is equivalent to the
ratio of the pseudorapidity asymmetry of negatively charged hadrons
relative to the asymmetry of the positively charged hadrons:
\begin{equation}
Y^{AB}_{Z}(p_T) = \left. Z^h_{AB}(p_T, \eta) \right|_{-\eta} 
 \left/ 
\left. Z^h_{AB}(p_T, \eta) \right|_{\eta} \right. \,\, .
\label{yzAB}
\end{equation}
\section{Results}
\label{res}

The gluon distribution is the least constrained, and therefore, 
constitute the
largest source of differences between the various global fits to
nuclear parton distributions (see \cite{Eskola:2009uj,Armesto:2006ph}). Thus the
differences in the predictions of the three sets under consideration
stem largely from the differences associated with their respective
gluon modifications. These differences are largest at the initial
factorization scale ($Q_{0}^{2} = 1.69,1.0$ GeV$^2$ for both EPS08 and
EPS09, and HKN07 respectively) and decreases with increasing $Q^2$ due
to evolution. With our choice of scales ($Q = Q_f =  p_T$) the results
presented here are sensitive to these differences up to the highest
$p_T$ considered, $p_T = 60$ GeV.  

Nuclear effects encoded in the nPDFs are $x$-dependent, and thus
different effects are present at different $x$: shadowing 
($x \lesssim 0.1$, depletion), antishadowing 
($0.1 \lesssim x \lesssim 0.3$, enhancement), EMC effect ($0.3 \lesssim 
x \lesssim 0.8$, depletion), and Fermi motion ($x > 0.1$, enhancement).
Due to the $x_b$-integration in Eq.~(\ref{eq:pdA}) 
($x_{min}^{B}(p_T,\eta) \le x_b \le 1$),
different nuclear effects are superimposed, thus it is
difficult to effect a puritanical isolation of these different nuclear
effects.

In the following subsections we present the results of our calculations for nuclear
modification factor, pseudorapidity asymmetry, and charge ratio for
the production of charged hadrons in dAu collisions at RHIC and pPb 
collisions at the LHC. It should be emphasized that the results from
both HKN07 and EPS09 nPDFs are from their respective central fits.  
%
\subsection{Nuclear modification factor of charged hadrons at RHIC and the LHC}
\label{rth}
\subsubsection{Charged hadron modification factor at RHIC}
\begin{figure}[!h]
\begin{center} 
\includegraphics[width=8.5cm, height=9.5cm, angle=270]{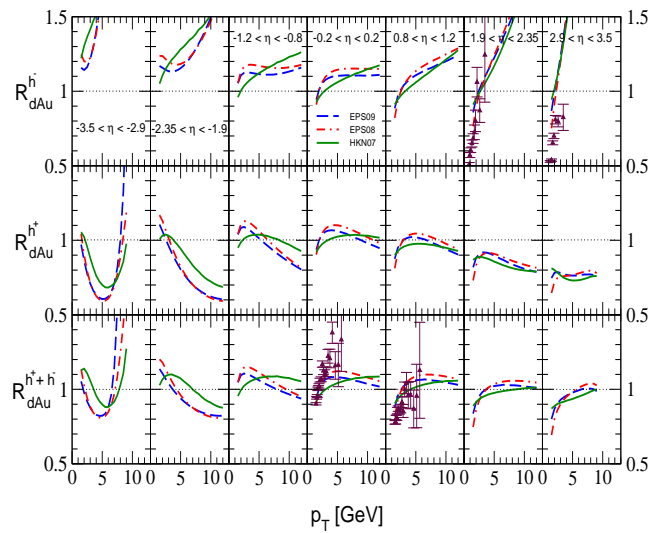}
\end{center}
\caption[...]{(Color Online) Nuclear modification factor,
$R_{dAu}$, for charged hadron production in d+Au collisions at RHIC
($\sqrt{s_{NN}} = 200$ GeV) from
backward ($\eta = -3.5$) to forward ($\eta = 3.5$) pseudorapidities. 
The solid line represents the HKN07 nPDFs, the dashed line the EPS09
nPDFs, while the dot-dashed line represents the 
EPS08 nPDFs. The filled triangles denote 
the BRAHMS data~\cite{Arsene:2004ux}.}
\label{fig:rdgrhic}
\end{figure}

 In Figure~\ref{fig:rdgrhic} we present our result for nuclear
 modification factor of charged hadrons at RHIC. Data exist for
 negatively charged hadrons at forward rapidities and for sum of
 charged hadrons around midrapidity, and the level of agreement of
 calculation with data is as shown in the figure. Our calculations
 extends over the positive and negative BRAHMS pseudorapidity intervals
 ($-3.5 < \eta < 3.5$), and thus cover reasonably well the whole
 spectrum of nuclear effects, viz, shadowing, antishadowing, EMC
 effects and Fermi motion. As is well known, large backward (negative) rapidities
 correspond to large Bjorken-$x$ momentum fractions in the gold
 nucleus while large forward (positive) rapidities correspond to small 
 Bjorken-$x$ momentum fractions. Thus as we move from backward to
 forward different nuclear effects come into play: from EMC/Fermi
 motion effects at large $x$ to progressively stronger suppression due
 to nuclear shadowing at very small Bjorken-$x$. A major limitation at
 RHIC is that the kinematics are such that the very small $x$ region ($x <<
 0.1$) is not appreciably accessed, except at very small $p_T$, where
 soft physics effects could be significant. This constraint is
 alleviated at LHC energies where it is possible to access the low $x$
 even at relatively high $p_T$.

A further complication is the presence of strong isospin effects on
the nuclear modification factor of charged hadrons in deuteron-gold 
collisions, in particular at very forward rapidities. 
This is readily apparent
from the appreciably different modification factors at forward rapidities
for positively charged and negatively charged hadrons respectively, in
contrast with the equality of these modification factors at forward
rapidities in proton-nucleus collisions. These effects are somewhat
minimized for the sum of charged hadrons, especially about midrapidity
($-1.2 < \eta < 1.2$), and so we therefore illustrate the influence
of nuclear effects on the nuclear modification factor of sum of charged hadrons.

Very backward rapidities ($-3.5 < \eta < -2.9$): at the lowest $p_T$ 
considered ($p_T \sim 1.5$), $x_{min}^{Au} \sim 0.2$, and thus the modification factor is
determined by the relative contributions from antishadowing, EMC
effect, and Fermi motion, leading to $R_{dAu} > 1$. As $p_T$
increases, $x_{min}^{Au}$ becomes larger and therefore only the EMC
effect and Fermi motion contribute, with $R_{dAu} < 1$. Finally, at
sufficiently high $p_T$, only the Fermi motion is present, thus
leading to an enhancement ($R_{dAu} > 1$).

Backward rapidities ($-2.35 < \eta < -0.8$): at very low $p_T$,
$x_{min}^{Au}$ is in the region of the advent of shadowing for $-2.35
< \eta < -1.9$, and already in the shadowing region for $-1.2
< \eta < -0.8$ respectively. Thus there is full contribution from
antishadowing (enhancement) for $-2.35 < \eta < -1.9$ and some
shadowing (suppression) for $-1.2 < \eta < -0.8$. At intermediate
$p_T$s there are contributions from both antishadowing and the
EMC/Fermi motion. For the interval $-2.35 < \eta < -1.9$ the
antishadowing contribution becomes negligible for $p_T > 5$ GeV, while
it is still present though small even at high $p_T$ for $-1.2 < \eta <
-0.8$. At high $p_T$ the modification factor is influenced dominantly
by the EMC effect.

Midrapidities ($-0.2 < \eta < 0.2$): shadowing contributes
appreciably at very low $p_T$ with negligible contributions at higher
$p_T$. The physics at intermediate and higher $p_T$ is determined by
the relative contributions from antishadowing (particularly for $p_T <
7.5$) and the EMC effect.

Forward rapidities ($0.8 < \eta < 2.35$): shadowing contributes
significantly at very low $p_T$ for $0.8 < \eta < 1.2$ and up till around
$p_T \sim 3$ GeV for $1.9 < \eta < 2.35$, although the effect is more
pronounced at lower $p_T$. The modification factor at
higher $p_T$ is determined by a mix of antishadowing and the EMC/Fermi
motion.

Very forward rapidities ($2.9 < \eta < 3.5$): here shadowing
dominates at very low $p_T$ and the effect persists up to $p_T \sim
4$ GeV. The effect of antishadowing is seen in the rise of $R_{dAu}$
between $4$ and $8$ GeV while the EMC effect is responsible for the
downward trend of $R_{dAu}$ at $p_T > 8$ GeV.

This simple analysis using the $x$-dependence of nuclear effects is of
course modified for positively and negatively charged hadrons
respectively. For negatively charged hadrons, the net effect is a
rapid enhancement of the modification factor with $p_T$, especially at
very backward and forward rapidities, and a less steep enhancement
with $p_T$ around midrapidities. The converse is the case for
positively charged hadrons, although the suppression is less steep
with increasing $p_T$ at both backward and forward rapidities. 

The observed predictions of the three different nPDFs sets reflect
the different relative strengths of the nuclear effects encoded in
these nPDFs, especially the nuclear gluon modifications. Even though
there are clear differences between the predictions of EPS09/EPS08
relative to those of HKN07 across board, a reflection of strong versus weak gluon
modifications, the distinction between EPS08 and EPS09 is not very
apparent at the very backward and very forward rapidities, except at
low $p_T$. In the interval $-2.35 < \eta < 2.35$, due to different 
gluon antishadowing and EMC, the distinction is more appreciable, 
despite the similarity in shape of the modification factor. 
The kinematics at the LHC will permit going deep
into the shadowing regime at forward rapidities, and thus one
anticipates clear-cut distinctions in the predictions of EPS08 
and EPS09 nPDFs for significant range of $p_T$.

We now compare our results with experimental data. The BRAHMS
Collaboration~\cite{Arsene:2004ux} has measured nuclear modification factor for sum of
charged hadrons in the intervals $-0.2 < \eta < 0.2$ and $0.8 < \eta <
1.2$, and for negatively charged hadrons in the intervals $1.9 < \eta
< 2.35$ and $2.9 < \eta < 3.5$. Figure~\ref{fig:rdgrhic} shows our
result for both negatively charged hadrons and sum of charged hadrons
for the relevant $\eta$ intervals. The effect of the different gluon
modifications present in the three nPDFs is apparent: EPS08
consistently has the strongest suppression at very low $p_T$, while
HKN07 has the weakest, for all pseudorapidity intervals. The effect of
gluon antishadowing and EMC is also readily seen for both $-0.2 < \eta <
0.2$ and $0.8 < \eta < 1.2$. In the interval $-0.2 < \eta < 0.2$,
EPS08, due to its stronger gluon modifications, has the best agreement
with data ($p_T < 6$ GeV) while HKN07 has the least. All three nPDFs
are in good agreement with data for both $0.8 < \eta < 1.2$ and 
$1.9 < \eta < 2.35$, although the reach of the data is quite limited for
the latter interval. 

At very forward rapidities ($2.9 < \eta < 3.5$) the data show a
suppression for the yield of negatively charged hadrons not adequately
reproduced by calculations for $p_T > 2$ GeV, albeit with the small
reach in $p_T$. This suppression has generated a lot of interest, with
diverse mechanisms being proffered to explain the phenomenon (see, for
instance, \cite{Kharzeev:2004yx,Guzey:2004zp,Hwa:2004in,Nemchik:2008xy}. 
Since our purpose
here is to compare predictions from different nPDFs, we will not go
into discussions of the various explanations of the observed
suppression. Some earlier studies of nuclear modification factor with
different nPDFs can be found in \cite{Guzey:2004zp,Vogt:2004hf,Levai:2006yd,Adeluyi:2008qk}.   
\subsubsection{Charged hadrons modification factor  at the LHC}

\begin{figure}[!h]
\begin{center} 
\includegraphics[width=8.5cm, height=9.5cm, angle=270]{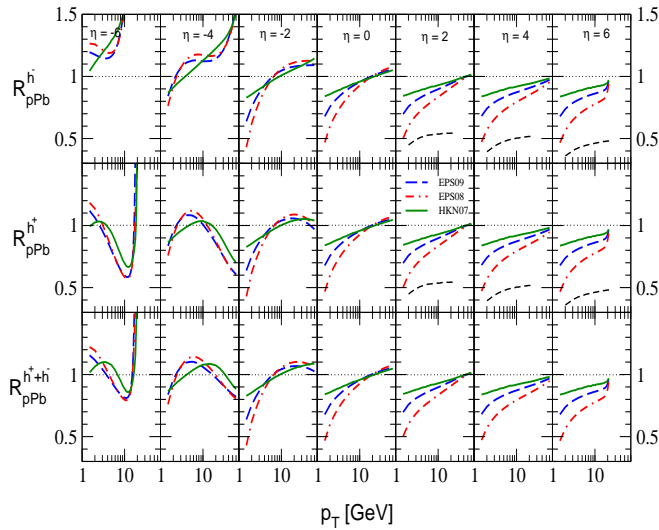}
\end{center}
\caption[...]{(Color Online) 
Nuclear modification factor,
$R_{pPb}$, for charged hadron production in p+Pb collisions at the
LHC ($\sqrt{s_{NN}} = 8.8$ TeV) from backward ($\eta = -6$) to forward
($\eta = 6$) pseudorapidities. 
The solid line represents the HKN07 nPDFs, the dashed line the EPS09
nPDFs, while the dot-dashed line represents the 
EPS08 nPDFs. 
The short dashed line is the result from a recent Color
Glass Condensate (CGC) calculation \cite{Albacete:2010bs}.
}
\label{fig:rpPblhc}
\end{figure}

We now consider nuclear modification factor of charged hadrons in 
proton-lead collisions at the LHC. Figure~\ref{fig:rpPblhc} shows our 
results for positively charged, negatively charged, and sum of charged
hadrons. Unlike what obtains in deuteron-gold collisions, the
modification factors for positively and negatively charged hadrons are
the same, except at very backward rapidities ($\eta = -6$ and $\eta =
-4$) where isospin effects in the Pb nucleus are appreciable, and the
comments above applicable. 
The behavior of the modification factor across the spectrum of
pseudorapidities mirrors the trend seen in d+Au collisions. We
illustrate this for the sum of charged hadrons at the very backward
rapidity.

At $\eta = -6$ and $p_T = 1.5$ GeV, $x_{min}^{Pb} \sim 0.07$ and thus 
there is negligible contribution from shadowing. The modification
factor thus reflects the competing contributions from
antishadowing/Fermi motion (enhancement) and the EMC effect
(suppression). Higher transverse momenta translate to larger
$x_{min}^{Pb}$, with less contribution from antishadowing, leading to  
$R_{pPb} < 1$. With $p_T > 10$ GeV, the dominant contribution is from
Fermi motion, with the attendant $R_{pPb} > 1$.

The major difference is the accessibility of the small-$x$ region even
at backward rapidities and at relatively high transverse momenta for
both midrapidities and forward rapidities. For instance, at $\eta =
-2$ and $p_T = 10$ GeV, $x_{min}^{Pb} \sim 0.01$, well within the
shadowing region while at $\eta = 2$ and $p_T = 60$ GeV, $x_{min}^{Pb}
\sim 0.001$, signifying a substantial contribution from shadowing at
this relatively high $p_T$. In fact, at very forward rapidities (
$\eta = 4, 6$) the influence of shadowing is so strong as to render 
$R_{pPb} < 1$ for all $p_T$ considered. The strong sensitivity of the
nuclear modification factor to gluon modifications enables a clear 
distinction of the predictions of the three nPDFs sets utilized in this study.

The Color Glass Condensate (CGC) is an attractive framework for 
addressing forward rapidity (small-$x$) physics (for a recent review see \cite{Gelis:2010nm}). 
For comparison purposes we show the predicted nuclear modification
factor for charged hadrons at rapidities $y = 2,4,6$ from a recent
calculation \cite{Albacete:2010bs} in this framework. Our result shows less
suppression and a more dramatic $p_T$ dependence than the CGC. 


\subsection{Pseudorapidity asymmetry of charged hadrons at RHIC and
  the LHC}
\label{nch}

We now address pseudorapidity asymmetry in d+Au collisions at RHIC and
p+Pb collisions at the LHC. A transparent way to understand the $p_T$
dependence of pseudorapidity asymmetry is to consider the asymmetry as
a ratio of backward and forward nuclear modification
factors~\cite{Adeluyi:2008gj}, the $pp$ cross section being symmetric with
respect to pseudorapidity. Thus it is rather straightforward to apply
the preceding analysis of the nuclear modification factor using the
$x$ dependence of nuclear effects to the $p_T$ dependence of
pseudorapidity asymmetry. Pseudorapidity asymmetry has been studied
for different asymmetric systems and at different energies in
\cite{Adeluyi:2008qk,Adeluyi:2008gj,Wang:2003vy,Barnafoldi:2005rb,Barnafoldi:2008rb} 
 
\subsubsection{Asymmetry at RHIC}

\begin{figure}[!h]
\begin{center} 
\includegraphics[width=8.5cm, height=9.5cm, angle=270]{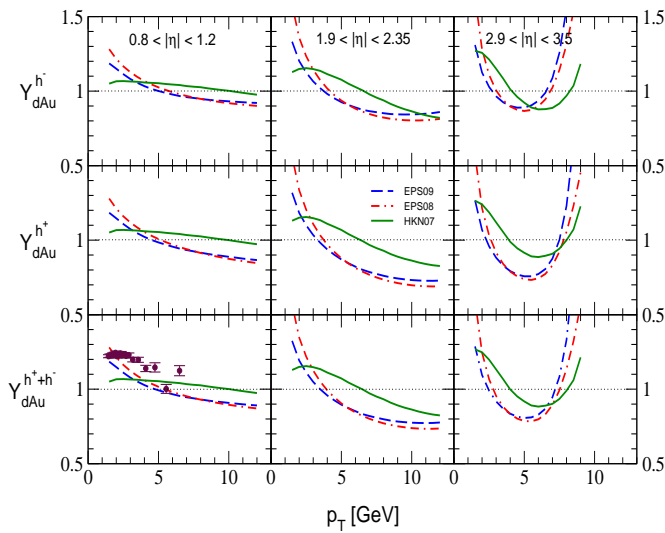}
\end{center}
\caption[...]{(Color Online) Pseudorapidity asymmetry,
$Y_{dAu}$ for charged hadrons at $0.8 < |\eta| < 1.2$,
$1.9 < |\eta| < 2.35$, and $2.9 < |\eta| < 3.5$ from d+Au collisions
at RHIC ($\sqrt{s_{NN}} = 200$ GeV). The solid line represents the
HKN07 nPDFs, the dashed line the EPS09 nPDFs while the dot-dashed line 
corresponds to the EPS08 nPDFs. The filled circles denote 
the STAR data~\cite{Abelev:2006pp} for $0.5 \le |\eta| \le 1.0$.}
\label{fig:asydAu}
\end{figure}

In Figure~\ref{fig:asydAu} we present the
pseudorapidity asymmetry for charged hadrons and sum of charged
hadrons produced in deuteron-gold collisions at RHIC. In the interval
$0.8 < |\eta| < 1.2$ the calculated asymmetry is practically the same
for both positively and negatively charged hadrons, and by
implication, the sum of charged hadrons. For $p_T < 6$ GeV both EPS09
and EPS08 yield a positive asymmetry ($> 1$), with the EPS08 giving a larger
asymmetry especially at low transverse momenta due to the stronger
gluon modifications. Above $p_T = 6$ GeV, the two nPDFs set predict a
negative asymmetry ($< 1$) of essentially the same magnitude. The
effect of a rather weak gluon modification in HKN07 is apparent: the
predicted asymmetry at low $p_T$ is less than that of both EPS08 and
EPS09. In fact, the predicted asymmetry is positive for all transverse
momentum considered, with the asymmetry essentially unity at large
$p_T$. The result for the sum of charged hadrons is compared with the
available experimental data from the STAR Collaboration
~\cite{Abelev:2006pp} for the interval $0.5 \le |\eta| \le 1.0$. 
Since this interval is
not far from midrapidity, the calculated asymmetry is rather weak, in
agreement with experimental observation \cite{Arsene:2004cn,Back:2004mr}. 

Larger asymmetries are expected as we move away from midrapidity. This
is readily apparent in the interval $1.9 < |\eta| < 2.35$ where the
calculated asymmetries for all three nPDFs sets are quite
significant. While HKN07 predicts essentially the same asymmetry for
both charge specie and the sum of charged hadrons, both EPS09 and
EPS08 predict, at higher transverse momenta, a larger negative
asymmetry for the positively charged hadrons relative to the
negatively charged hadrons. The effect of the different nuclear gluon
modifications in the three nPDFs is also apparent: EPS08 yields the
largest positive (negative) asymmetry at low (high) $p_T$ while HKN07
yields the smallest, especially for positively charged hadrons and sum
of charged hadrons. Also there are small but discernible differences between
EPS09 and EPS08 for all transverse momenta, unlike the case for the
interval $0.8 < |\eta| < 1.2$ where both predict the same
asymmetry for $p_T > 6$ GeV. Whereas both EPS09 and EPS08 yield
positive asymmetry for $p_T < 4$ GeV and negative for $p_T > 4$ GeV,
HKN07 gives negative asymmetry only for  $p_T > 7$ GeV.

The above observations are also relevant for the asymmetries from the
three sets in the interval $2.9 < |\eta| < 3.5$. The difference
between EPS09 and EPS08, though small, is quite distinct for all
charged species and at all relevant $p_T$. Both predict more
significant asymmetry for positively charged hadrons than for
negatively charged hadrons and for sum of charged hadrons. The physics
of the asymmetry (a ratio) in this interval is governed largely by
shadowing/antishadowing (denominator) and EMC/Fermi motion
(numerator). Thus both EPS08 and EPS09 predict negative asymmetry for
$2.8 < p_T < 7.5$ GeV and positive everywhere else. On the other hand,
HKN07 yields a positive asymmetry for $p_T < 4.5$ GeV and  $p_T < 8.5$
GeV, with negative asymmetry for transverse momenta within these
bounds.

It is pertinent to remark at this stage that even though the
predictions from EPS08/EPS09 on the one hand, and those of HKN07 on the
other hand are clearly distinguishable, the kinematic reach at RHIC is
such as to limit the clarity of the distinction between a strong gluon
modification scenario as present in EPS08 and the somewhat more moderate
gluon modification as in EPS09. The kinematics at the LHC allow such a
distinction, and thus the study of pseudorapidity asymmetry at the LHC
may shed some light on the actual strength of nuclear gluon modifications.

\subsubsection{Asymmetry at the LHC}

\begin{figure}[!h]
\begin{center} 
\includegraphics[width=8.5cm, height=9.5cm, angle=270]{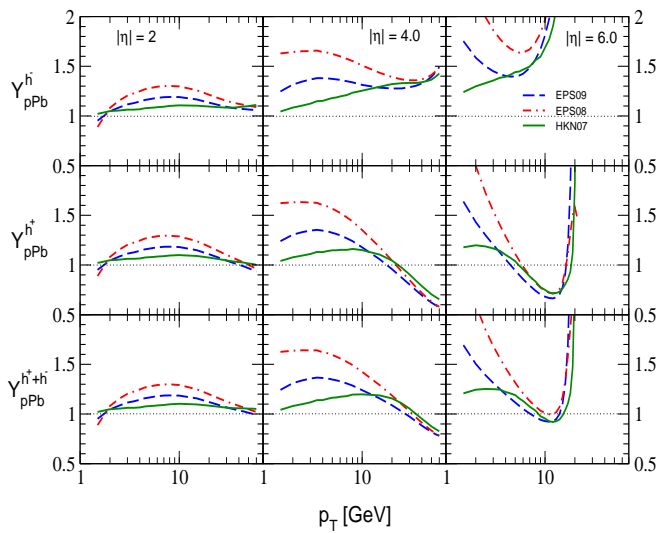}
\end{center}
\caption[...]{(Color Online) Pseudorapidity asymmetry,
$Y_{pPb}$ for charged hadrons at $|\eta| = 2$,
$|\eta| = 4$, and $|\eta| = 6$ from p+Pb collisions
at the LHC ($\sqrt{s_{NN}} = 8.8$ TeV). The solid line represents the
HKN07 nPDFs, the dashed line the EPS09 nPDFs while the dot-dashed line 
corresponds to the EPS08 nPDFs.}
\label{fig:asypPb}
\end{figure}

Figure~\ref{fig:asypPb} depicts our result for pseudorapidity
asymmetry of charged hadrons at the LHC. Let us consider the first
interval: $|\eta| = 2$. Even at this relatively small pseudorapidity
the different predictions of the three nPDFs are clearly
distinguishable. Both EPS09 and EPS08 predict a negative asymmetry for
$p_T < 2$ GeV, with the EPS08 yielding about twice the value of EPS09
at the lowest $p_T$ considered. Thereafter both predict a positive
asymmetry which peaks around $p_T = 9$ GeV and then decreases towards
unity at very high $p_T$. At this peak, EPS08 yields an asymmetry of
about $30\%$ while EPS09 yields about $20\%$. The asymmetry predicted
by HKN07 is characteristically different: it rises approximately 
monotonically as $p_T$ increases and is positive for all $p_T$
considered. At the highest $p_T$ considered ($p_T = 60$ GeV), the
asymmetry predicted by HKN07 is roughly the same as EPS08 for both
negatively charged hadrons and sum of charged hadrons, and about
$15\%$ greater for positively charged hadrons. From Figure~\ref{fig:asypPb}
it is apparent that the region $3 < p_T < 20$ GeV gives the best
discriminatory ability for the considered pseudorapidity.

The distinguishability of the different predictions of pseudorapidity
asymmetry by the three nuclear parton distributions sets is
appreciably magnified for the interval $|\eta| = 4$, especially for
$p_T < 20$ GeV. In this $p_T$ region the predicted asymmetries are
positive and approximately independent of the charge specie, with
EPS08 giving the largest asymmetry and HKN07 the smallest. Above this
region the predicted asymmetries are positive for negatively charged
hadrons and gradually turn negative and close in magnitude as $p_T$ 
increases for positively charged hadrons and sum of charged hadrons. 
Thus the region $p_T < 15$ GeV affords the best opportunity to discriminate
between the different scenarios of gluon modifications as enshrined in
the three nPDFs. 

The above observations for $|\eta| = 4$ apply equally well for the
pseudorapidity asymmetry in the interval $|\eta| = 6$. Here, due to
phase space limitations, good discriminatory ability is limited to the
transverse momentum region $p_T < 10$ GeV. While the predicted
asymmetries remain positive for all $p_T$ in the case of negatively
charged hadrons, positively charged hadrons (sum of charged
hadrons) exhibit a crossover from positive asymmetry to negative
asymmetry around $p_T = 6$ GeV ($p_T = 10$ GeV) and a sharp rise back
to positive asymmetry at around $p_T = 15$ GeV.

Thus contrary to what obtains at RHIC energies, pseudorapidity
asymmetry at the LHC offers the potential of clearly distinguishing
between various nPDFs with different nuclear gluon modifications.

\subsection{Charge ratio and its asymmetry at RHIC and the LHC}
\label{pch} 

\begin{figure}[!h]
\begin{center} 
\includegraphics[width=8.5cm, height=9.5cm, angle=270]{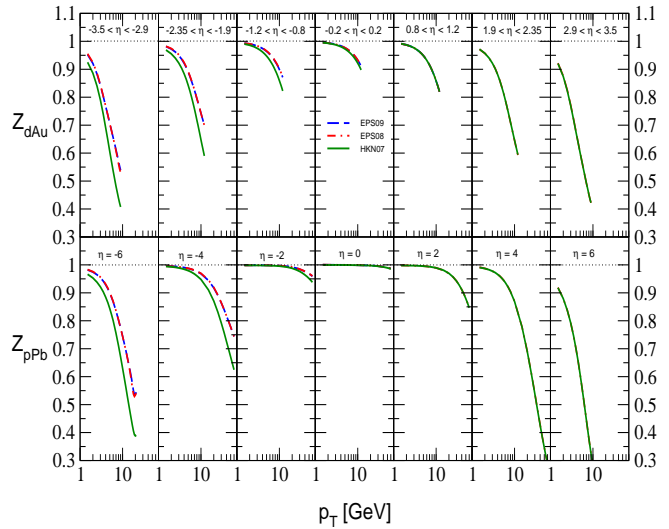}
\end{center}
\caption[...]{(Color Online) Top panel: Charge ratio,
$Z_{dAu}$, for charged hadron production in d+Au collisions at RHIC
($\sqrt{s_{NN}} = 200$ GeV) from
backward ($\eta = -3.5$) to forward ($\eta = 3.5$) pseudorapidities.
Bottom panel: Charge ratio,
$Z_{pPb}$, for charged hadron production in p+Pb collisions at the
LHC ($\sqrt{s_{NN}} = 8.8$ TeV) from backward ($\eta = -6$) to forward ($\eta = 6$) pseudorapidities. 
In general solid line represents the HKN07 nPDFs, dashed line the EPS09
nPDFs, while dot-dashed line represents the EPS08 nPDFs.}
\label{fig:chrdAupPb}
\end{figure}

The top panel of Figure~\ref{fig:chrdAupPb} shows the result of our
calculation of the charge ratio in d+Au collisions at RHIC. 
The ratio is consistently below unity for all $\eta$ intervals and transverse
momenta considered, with the implication of a suppression of
negatively charged hadrons relative to the positively charged
hadrons, in accord with expectations from fragmentation. 
The suppression is smallest around midrapidity where the
charge ratio is close to unity, and becomes progressively more
enhanced as one moves towards larger negative and positive rapidities.
In the very
backward rapidity interval considered ($-3.5 < \eta < -2.9$), the
sensitivity of the charge ratio to nPDFs is most pronounced. While
both EPS09 and EPS08 manifest practically the same behavior, HKN07 
displays a greater suppression than both EPS09 and EPS08. This
difference decreases significantly as we move towards positive
rapidities, such that already at the interval $0.8 < \eta < 1.2$ all
three nPDFs exhibit the same behavior for the charge ratio. Thus at
very forward rapidities the charge ratio is independent of the
differences in the nuclear
parton distributions, and is thus solely determined by the
fragmentation functions.

The same trend is observed for p+Pb collisions at the LHC. Both EPS09
and EPS08 predict essentially the same behavior for all rapidities and
transverse momenta. As in the case of d+Au collisions at RHIC, the
sensitivity is most pronounced for the very backward pseudorapidities
and disappears at very forward rapidities. At midrapidity
($\eta = 0$), all three nPDFs yield a ratio that is essentially
unity. The HKN07 nPDFs set manifests quantitatively a greater
suppression at very backward rapidities, same as exhibited in d+Au
collisions at RHIC. 

We now present the result for the pseudorapidity asymmetry of the
charge ratio, $Y^{AB}_{z}$. As remarked above, this is just the ratio of the
usual pseudorapidity asymmetry of negative hadrons relative to the
positively charged hadrons. Since the charge ratio is close to unity
around midrapidity, the focus of interest is the behavior at both very
forward (small (large) $x$ in target (projectile)) and backward 
(large (small) $x$ in target (projectile))
rapidities. Thus the pseudorapidity asymmetry of the charge ratio
helps to display the degree of asymmetry of the suppression relative
to midrapidity. 

\begin{figure}[!h]
\begin{center} 
\includegraphics[width=8.5cm, height=9.5cm, angle=270]{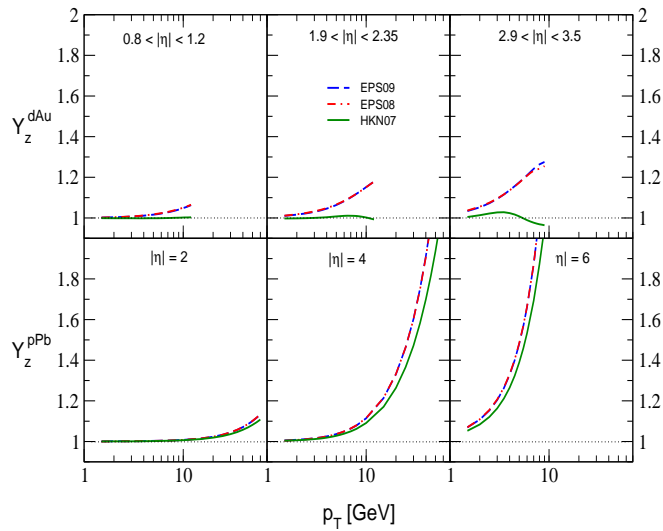}
\end{center}
\caption[...]{(Color Online) 
Top panel: Pseudorapidity asymmetry of charge ratio,
$Y^{dAu}_{z}$ for charged hadrons at $0.8 < |\eta| < 1.2$,
$1.9 < |\eta| < 2.35$, and $2.9 < |\eta| < 3.5$ from d+Au collisions
at RHIC ($\sqrt{s_{NN}} = 200$ GeV).
Bottom panel: Pseudorapidity asymmetry,
$Y^{pPb}_{z}$ for charged hadrons at $|\eta| = 2$,
$|\eta| = 4$, and $|\eta| = 6$ from p+Pb collisions
at the LHC ($\sqrt{s_{NN}} = 8.8$ TeV). 
In both cases the solid line represents the HKN07 nPDFs, the dashed line the EPS09
nPDFs, while the dot-dashed line represents the EPS08 nPDFs.}
\label{fig:asychrdAupPb}
\end{figure}

In Figure~\ref{fig:asychrdAupPb} we display the asymmetry for d+Au
collisions (top panel) and p+Pb collisons (bottom panel). For
deuteron-gold collisions, the asymmetry is positive and increasing 
with increasing $\eta$ for both EPS09 and EPS08. In the case of HKN07,
the asymmetry is consistent with unity for the first two $\eta$
intervals, while it is slightly negative at large $p_T$ for 
$2.9 < |\eta| < 3.5$.
Let us now consider p+Pb collisions at the LHC. Both EPS09 and EPS08
predict positive asymmetry for all rapidities considered, with the
asymmetry increasing sharply with increasing $\eta$. This trend is
replicated by the HKN07 set. The degree of asymmetry is consistent for
all three sets at $|\eta| = 2$, with slight differences at high $p_T$
for larger rapidities.

\section{Conclusion}
\label{concl}

In this study we have investigated the effects of different strengths
of nuclear gluon modifications on charged hadron production at both 
RHIC and the LHC, using three recent nuclear parton distributions. We
considered asymmetric light-on-heavy systems where final-state effects 
(apart from fragmentations leading to final charged hadrons) are
absent. The present treatment include observables like nuclear
modification factor, pseudorapidity asymmetry, and charge ratio for
both negative and positive pseudorapidites and up to relatively high
transverse momenta, thereby ensuring adequate coverage of all relevant
nuclear effects encoded in the nuclear parton distributions. 

Each of the observables considered has its own strengths and weaknesses
as a probe of nuclear effects in nuclear parton distributions. Nuclear
modification factor probes directly nuclear effects for any $p_T$ and
at any pseudorapidity $\eta$. Its drawback is the dependence on knowledge of 
the requisite $pp$ yield. The associated modification factor,
central-to-peripheral ratio, while not dependent on the $pp$ cross
section, is geometry dependent. The pseudorapidity asymmetry, on the
other hand, is independent of the $pp$ cross section, but requires the
cross section of p(d)A collisions at both forward and backward
rapidities. It is thus somewhat less "clean" than the nuclear
modification factor, but still has good potentials for constraining
nPDFs. As shown above, the charge ratio is only sensitive in the
backward region (large $x$ in Au(Pb)); so it may offer some
discriminating abilities for nuclear effects at large $x$.

While there are some clear differences between the predictions of the three
nPDFs for the nuclear modification factor and pseudorapidity asymmetry
at RHIC, the distinctions are significantly enhanced at the LHC, even
up to relatively high transverse momenta. In particular, at very
forward rapidites where nuclear shadowing is expected to dominate, the
predictions of the nuclear modification factor for all three nPDFs are
directly correlated with the strengths of their gluon
modifications. This is also true for the predicted pseudorapidity
asymmetries.

In conclusion, asymmetric p(d)A collisions, especially at the LHC,
have great potential not only to shed light on the underlying dynamics
of relativistic nuclear collisions, but also as a source of data to
help better constrain nuclear gluon distributions.

\section{Acknowledgments}
\label{ack}
This work is supported in part by the NSF grant PHY0757839.
%
%

\end{document}